# Soft X-ray Absorption Spectroscopy and Magnetic Circular Dichroism as *Operando* Probes of Complex Oxide Electrolyte Gate Transistors

Biqiong Yu,[1] Guichuan Yu,[1] Jeff Walter,[2,3] Vipul Chaturvedi,[2] Joseph Gotchnik,[2] John W. Freeland,[4] Chris Leighton,[2]* Martin Greven[1]*

[1]School of Physics and Astronomy, University of Minnesota, Minneapolis, MN 55455, USA
[2]Department of Chemical Engineering and Materials Science,
University of Minnesota, Minneapolis, MN 55455, USA
[3]Department of Physics, Augsburg University, Minneapolis, MN 55454, USA
[4]Advanced Photon Source, Argonne National Laboratory, Argonne, IL 60439

**Abstract:** Electrolyte-based transistors utilizing ionic liquids/gels have been highly successful in the study of charge-density-controlled phenomena, particularly in oxides. Experimental probes beyond transport have played a significant role, despite challenges to their application in electric double-layer transistors. Here, we demonstrate application of synchrotron soft X-ray absorption spectroscopy (XAS) and X-ray magnetic circular dichroism (XMCD) as *operando* probes of the charge state and magnetism in ion-gel-gated ferromagnetic perovskite films. Electrochemical response *via* oxygen vacancies at positive gate bias in $LaAlO_3(001)/La_{0.5}Sr_{0.5}CoO_{3-\delta}$ is used as a test case. XAS/XMCD measurements of 4-25 unit-cell-thick films first probe the evolution of hole doping (from the O *K*-edge pre-peak) and ferromagnetism (at the Co *L*-edges), to establish a baseline. *Operando* soft XAS/XMCD of electrolyte-gated films is then demonstrated, using optimized spin-coated gels with thickness ~1 μm, and specific composition. Application of gate voltages up to +4 V is shown to dramatically suppress the O *K*-edge XAS pre-peak intensity and Co *L*-edge XMCD, thus enabling the Co valence and ferromagnetism to be tracked upon gate-induced reduction. Soft XAS and XMCD, with appropriate electrolyte design, are thus established as viable for the *operando* characterization of electrolyte-gated oxides.

*Corresponding authors: leighton@umn.edu, greven@umn.edu

Electric double-layer transistors (EDLTs) employing ionic liquids/gels have proven successful in the study of charge-density-dependent effects in many materials[1–4]. This approach allows the density of doped carriers to be varied in a single device, to extremely high charge densities (> $10^{14}$ cm$^{-2}$)[1-14], at least an order of magnitude above conventional (e.g., SiO$_2$-based) field-effect devices. EDLTs are thus widely used to study electronic and magnetic phase transitions[1-18] and map phase diagrams[11,16]. For instance, electrically-induced superconductivity was observed in electrolyte-gated KTaO$_3$[8], SrTiO$_3$[14], and high-temperature superconductors[2,10,16–18]. Control of the insulator-metal transition was achieved in VO$_2$[12,19–21] and nickelates[7,13], and modulation of magnetism was found possible in Ti$_{1-x}$Co$_x$O$_2$[9], La$_{1-x}$Sr$_x$MnO$_3$[6,22], and La$_{1-x}$Sr$_x$CoO$_{3-\delta}$[23–25].

It is increasingly recognized that EDLT operating mechanisms are typically not purely electrostatic, and that there exists a range of alternative responses[4,19,21,23–28]. In oxides, for example, the gate response can be electrochemical, where the EDL electric field induces redox chemistry, often *via* oxygen vacancy (V$_O$) formation/annihilation. Evidence for this originally came from transport measurements in controlled atmosphere, with tracking of irreversibility[21,23,29–33]. Recently, however, synchrotron-based hard X-ray diffraction[24,34–36] and hard XAS[34,37], as well as neutron reflectometry[24,25] have been developed as *operando* probes of EDLTs, complementing transport. Our recent *operando* X-ray diffraction on ion-gel-gated La$_{0.5}$Sr$_{0.5}$CoO$_{3-\delta}$ (LSCO) highlighted the importance of gate-bias polarity: a large lattice expansion was found at positive gate bias ($V_g$) due to V$_O$ formation, whereas only minor structural changes occurred at negative $V_g$[24]. Importantly, the $V_g$-induced V$_O$ penetrated the entire film thickness, confirmed by depth-sensitive neutron reflectometry; this is enabled by the high diffusivity of V$_O$ in LSCO, which focuses much attention on redox control of these compounds[23–25,36,38,39].

Although application of EDLTs is growing, an element-sensitive *operando* probe such as XAS, which enables element-specific determination of valence, is not fully developed. Some absorption spectroscopy studies in the hard X-ray regime[34,37] provided insights into EDLT gating mechanisms, but *soft* XAS at the O $K$ and transition-metal $L$ edges, which directly probes transition-metal electronic/magnetic structure, has not yet been performed in an *operando* fashion. The main challenges in such *operando* measurements (as opposed to *ex situ* studies of non-volatile gate effects) include penetrating thick electrolyte layers with soft X-rays and getting absorption signals out.

Here, we establish soft XAS and XMCD, its magnetic variant, as *operando* probes of charge state and magnetism in electrolyte-gated (specifically ion-gel-gated) epitaxial LSCO films. Baseline information is first gathered from 4-25 unit-cell-thick bare films by determining the evolution in hole doping from O $K$ edge XAS and in ferromagnetism from Co $L$ edge XMCD. The ion gel (i.e., ionic liquid in a polymer network) used for *operando* measurements was optimized in terms of thickness (to ~1 μm) and composition. We show that, upon applying $V_g$ = +4 V, the O $K$ edge XAS pre-peak intensity is strongly suppressed, indicating a substantial decrease in effective hole density. Concomitantly, the Co magnetic moment is substantially reduced, as revealed in energy- and magnetic-field-dependent XMCD. These results, which complement prior X-ray diffraction and neutron reflectometry[23,24], not only yield new insight into gating mechanisms in oxide EDLTs, but also demonstrate a powerful *element-sensitive* approach to *operando* studies of gated oxides.

As described previously[23], epitaxial LSCO films were grown using high-pressure-oxygen reactive sputtering in 1.4 Torr of $O_2$ from ceramic LSCO targets onto 10 × 10 mm² LaAlO$_3$(001) substrates at 600 °C. Extensive structural, chemical, magnetic and transport characterization has been published[40–43]. To prepare EDLTs, 28-unit-cell-thick films were patterned into 3 × 3 mm² channels between two Pt side-gate electrodes[23,24]. Ion gels based on the ionic liquids 1-ethyl-3-methylimidazolium bis (trifluoro-methylsulfonyl) imide (EMI:TFSI) or 1-ethyl-3-methylimidazolium dicyanamide (EMI:DCA) in poly(vinylidene fluoride-*co*-hexafluoropropylene) (P(VDF-HFP)) were spin-coated across the channel and gate electrodes to complete devices. EMI:TFSI was chosen for consistency with prior work, whereas EMI:DCA was chosen as it is oxygen-free and thus may avoid contamination of the O *K* edge. Devices were then wired up and immediately loaded into the beamline vacuum chamber (kept at $<10^{-8}$ Torr).

Soft synchrotron X-ray measurements were performed at beamline 4-ID-C of the Advanced Photon Source. Three different detectors/measurement modes are available: total-electron-yield mode (TEY), total-fluorescence-yield mode (TFY), and reflectivity mode (REF). TEY is dominated by the photoelectric effect, and its probing depth is thus limited to the top unit cells. TFY measures emitted fluorescence photons, and thus has a larger probing depth due to the longer photon mean-free-path. REF deals with the reflection of the X-ray beam by the sample/device at a specific angle, with intensity dependent on the real and imaginary parts of the refractive index. Based on this, bare films were measured in TEY and TFY modes, in grazing-incidence geometry. For *operando* gating, on the other hand, in order to utilize both TFY and REF modes, the films were rotated by 30° facing the X-ray beam, with the fluorescence and reflectivity detectors at 90° and 60°, respectively (Fig. 3a). TEY mode was not used in gating experiments, as the overlying ion gel inhibits release of photoelectrons. The incident polarization was switched between left-circular and right-circular during the measurements. The sum ($I^+ + I^-$) of these signals probes the electronic environment of the electrons (XAS), whereas the difference ($I^+ - I^-$) contains magnetic information (XMCD). The EDLTs were operated at $<10^{-8}$ Torr, with $V_g$ applied at 310 K for 30 min, with *in situ* monitoring of electronic transport. *Operando* XAS and XMCD were measured at 75 K (below the Curie temperature, $T_C \approx 220$ K at this thickness), with a magnetic field $H = 4,000$ Oe along the X-ray beam; 75 K was chosen so that a large magnetic moment could be reached while keeping the saturation field safely below the maximum available (4,600 Oe).

We first gathered baseline information for LaAlO$_3$(001)/LSCO films by probing the evolution of hole doping and magnetism in bare films. Fig. 1 shows the O *K*- and Co *L*-edge XAS and XMCD spectra (TFY) for films of 4, 8, and 25 unit cells (u.c.). The data are normalized to 1 at the main peaks (540 eV for the O *K* edge, 780 eV for the Co *L* edge). The O *K* edge pre-peak near 527 eV is seen in all films (Fig. 1a), its intensity noticeably decreasing with decreasing thickness. This pre-peak has been well characterized in bulk LSCO and linked to the O 2*p* hole density[44]. The clear pre-peak decrease with decreasing thickness thus indicates an obvious decrease in effective hole doping in thinner films. Earlier work tied this to an increase in O deficiency δ near the substrate, leading to a decrease in effective hole doping, i.e., $x_{eff} = x - 2\delta$, in the simplest model[41]. This O deficiency was in turn linked to V$_O$ ordering, which was shown to be the lattice mismatch accommodation mechanism in this system[42]. This effect is also seen in the Co *L*-edge XAS (Fig. 1b), from the ~0.6 eV shift to lower energy; this is also known to indicate a decrease in Co valence[45]. Consequently, the Co magnetic moment is reduced on going from 25 to 8 u.c. thickness. In particular, as seen from the Co *L*-edge TFY XMCD in Fig. 1d, we find no evidence for ferromagnetism in the 4 u.c. film, whereas 8 and 25 u.c. films are clearly ferromagnetic. In bulk LSCO it is known that O holes also form magnetic states at the Fermi level, and that the O moment grows with doping [44]. We indeed observe an evolution of magnetism at O hole sites in TEY mode (Fig. 1c inset), although the low signal-to-noise ratio in TFY mode (Fig. 1c) results in an inability to resolve this signal. XMCD spectra at the O *K* edge were not measured further, since, as noted, the TEY mode was not possible in gating experiments.

A pivotal challenge in *operando* soft XAS and XMCD measurements is to optimize the thickness of the ion gel overlying the gated films: while a thin gel is desirable because of the low X-ray penetration depth (~1 μm at the Co $L$ edge), the gel must be thick enough to function electrically and achieve uniform gating. Optimization of ion gels for *operando* soft XAS/XMCD was achieved *via* a series of spin-coating experiments on Si/SiO$_x$ substrates. Solutions with varied polymer : ionic liquid : acetone(solvent) ratio (by mass) were prepared, heated to ~35 °C, and spun at 1500 rpm for 30 s. Tilt-view (45° with respect to the substrate plane) secondary electron scanning electron microscopy (SEM) images were then collected in a JEOL JSM-6010 PLUS/LA microscope, using 5 kV accelerating voltage. Images collected near an intentional scratch, so that the ion gel thickness (*d*) could be extracted, are shown in Fig. 2. Panels (a) – (d) and (e) – (h), for EMI:TFSI and EMI:DCA gels, respectively, show that ion gels become thinner with increasing ratios of ionic liquid and/or solvent. Excessively high ionic liquid/solvent ratios result in an "island and hole" morphology (Figs. 2d and 2h), unsuitable for gating. We thus established the optimized solutions to be in the range 1:150:50 for EMI:TFSI (Fig. 2c) and 1:350:50 for EMI:DCA (Fig. 2g), resulting in $d \approx 1.5 - 2.5$ μm. Upon testing optimized EMI:DCA gels, however (in fact, any EMI:DCA gels), we observed that the LSCO resistance increased by orders of magnitude in a matter of minutes following spin coating, whereas minimal change (< 0.1 %/min, consistent with prior work[23]) was observed for EMI:TFSI. The rest of the work presented here will thus utilize the 1:150:50 EMI:TFSI gel.

With ion gels optimized, we performed *operando* XAS/XMCD measurements of an electrolyte-gated LSCO film (28 u.c. thickness). Fig. 3a shows a schematic of the experimental setup. Throughout this study, only positive biases were applied, as a model test case. As noted, this $V_g$ polarity results in V$_O$ formation through the entire volume of LSCO films in this thickness regime, resulting in large resistivity[23] and lattice parameter increases[24], and a concomitant decrease in magnetization and Curie temperature. Figs. 3b and 3d show the O $K$ edge electronic structure changes at 75 K, after applying $V_g$ = +4 V at 310 K. Most noticeably, the 527 eV O 2$p$ pre-peak is seen to be essentially extinguished at $V_g$ = +4 V, in both TFY (Fig. 3b) and REF (Fig. 3d, inset) modes. This suppression is consistent with formation of a high density of $V_O$ at positive biases, decreasing the effective hole concentration ($x_{eff} = x - 2\delta$) by compensation of doped holes. Consistently, the Co $L_3$ edge XAS peak shifts ~1.2 eV to lower energy at $V_g$ = +4 V (Fig. 3c), indicating decreasing Co formal valence. The responses of the O $K$ edge and Co $L$ edge spectra to *operando* gating are thus qualitatively consistent. We note that, in addition to the LSCO films, both the LAO substrate and the ion gel include oxygen. The slight spectral changes at the O $K$ edge above 530 eV might thus involve not only LSCO, but also the ion gel, e.g., through beam damage.

Subsequent gate-induced magnetism changes were probed via Co $L_3$-edge XMCD. Figs. 4a and 4b show the energy dependence of the TFY and REF XMCD, respectively; these data were obtained at 75 K with an effective magnetic field in the film plane of ~3,500 Oe. At $V_g$ = 0, a peak occurs around 780.5 eV in TFY mode, whereas two peaks are seen in REF mode due to interference effects, indicative of a substantial Co moment, as expected. Upon application of $V_g$ = +4 V these peaks essentially vanish, meaning that the Co moment is strongly suppressed by the hole doping decrease due to $V_O$ formation, as evidenced by XAS. Fig. 4c shows corresponding hysteresis loops measured at the peak energy (~780.5 eV). The loops show pronounced magnetization and hysteresis (1,740 Oe coercivity) at $V_g$ = 0. At $V_g$ = +4 V, however, the saturation magnetization decreases substantially (by a factor of ~7), in agreement with prior neutron reflectometry measurements[24]. In the latter, a decrease of peak magnetization by a factor of ~8 was observed at 30 K for $V_g = +3$ V; our *operando* XMCD results are thus in quantitative agreement with prior neutron reflectometry[24].

In summary, we have demonstrated *operando* soft XAS/XMCD measurements on electrolyte-gated oxides, using ion-gel-gated LSCO films as a test case. Baseline information regarding hole doping and ferromagnetism was first established by measuring O *K*- and Co *L*-edge XAS/XMCD on bare LAO/LSCO films with thickness 4, 8, and 25 u.c. To overcome the penetration depth problem, the ion-gel was optimized with regard to composition and thickness, the latter reaching ~1 μm. Data for gated films were then obtained in TFY and REF modes. Application of $V_g = +4$ V resulted in dramatic suppression of the O *K* edge XAS pre-peak, indicating a significant decrease in hole doping due to $V_O$ formation at positive $V_g$. Concomitantly, a significantly reduced Co moment was observed in energy- and field-dependent XMCD spectra. Our investigation of LSCO-based ion-gel EDLTs via *operando* soft XAS/XMCD therefore provides direct evidence for electrochemical changes in hole doping and magnetism. This lays the foundation for *operando* soft XAS/XMCD studies of other electrolyte-gated oxides, also relevant to *operando* studies of battery, ionic conductor, and supercapacitor materials.

**Acknowledgements:** This work was primarily supported by the National Science Foundation through the UMN MRSEC under DMR-1420013. Parts of this work were carried out in the Characterization Facility, UMN, which receives partial support from NSF through the MRSEC program. Portions of this work were also conducted in the Minnesota Nano Center, which is supported by the National Science Foundation through the National Nano Coordinated Infrastructure Network, Award Number NNCI -1542202. This research used resources of the Advanced Photon Source, a DOE Office of Science User Facility operated by Argonne National Laboratory under Contract No. DE-AC02-06CH11357.

**FIGURE CAPTIONS**

**Fig. 1.** (a) O $K$-edge and (b) Co $L$-edge XAS spectra of LaAlO$_3$(001)/La$_{0.5}$Sr$_{0.5}$CoO$_{3-\delta}$ films with thickness ($t$) 4, 8, and 25 unit cells (u.c.). (c) O $K$-edge and (d) Co $L$-edge XMCD spectra of the same films. All data taken with grazing incidence X-rays in TFY mode, at temperature $T$ = 30 K, with a field $H$ = 4,600 Oe along the incident beam (i.e., the full field was in the film plane). In (b), a shift from 25 to 8 u.c. is observed (about 0.6 eV at the $L_3$ edge, as indicated). Inset to (c): TEY XMCD spectra near the pre-peak.

**Fig. 2.** Tilt-view (45°) secondary electron scanning electron microscopy images of ion gels spin-coated on Si/SiO$_x$ substrates at different ratios (by mass) of PVDF-HFP : ionic liquid : acetone, as shown. The ionic liquid was (a-d) EMI:TFSI and (e-h) EMI:DCA. Each coated film was scratched with a blade to enable thickness determination, with the resulting value ($d$) shown in each panel.

**Fig. 3.** (a) Schematic of the device and experimental setup for *operando* XAS/XMCD measurements of ion-gel-gated La$_{0.5}$Sr$_{0.5}$CoO$_{3-\delta}$ (LSCO) films. The orange area is the LSCO film (28 unit-cells-thick), and the grey pads are Pt electrodes for bias voltage application and *operando* transport measurements. The sample is rotated by $\theta$ = 30° such that both reflectivity and fluorescence channels can be probed. For XMCD (Fig. 4), a field of $H$ = 4,000 Oe was applied along the X-ray beam, resulting in a field of about 3,500 Oe in the film plane. The X-ray beam was defocused to cover the entire sample and minimize beam damage. Gate bias ($V_g$)-dependent XAS taken in (b) REF mode at the O $K$ edge, (c) TFY mode at the Co $L_3$ edge and (d) TFY mode at the O $K$ edge. '*' denotes the pre-peak around 527 eV, as highlighted in the inset to (d). All data taken at $T$ = 75 K.

**Fig. 4.** Gate-bias voltage ($V_g$) dependence of the ferromagnetic properties of 28-unit-cell-thick La$_{0.5}$Sr$_{0.5}$CoO$_{3-\delta}$ probed by XMCD at the Co $L_3$ edge in (a) TFY and (b) REF modes. Data taken

at $T = 75$ K with $H = 4,000$ Oe along the X-ray beam. (c) Change of XMCD hysteresis loops with $V_g$, as measured in REF XMCD mode at 75 K with phonon energy 780.5 eV. Lines are guides to the eye.

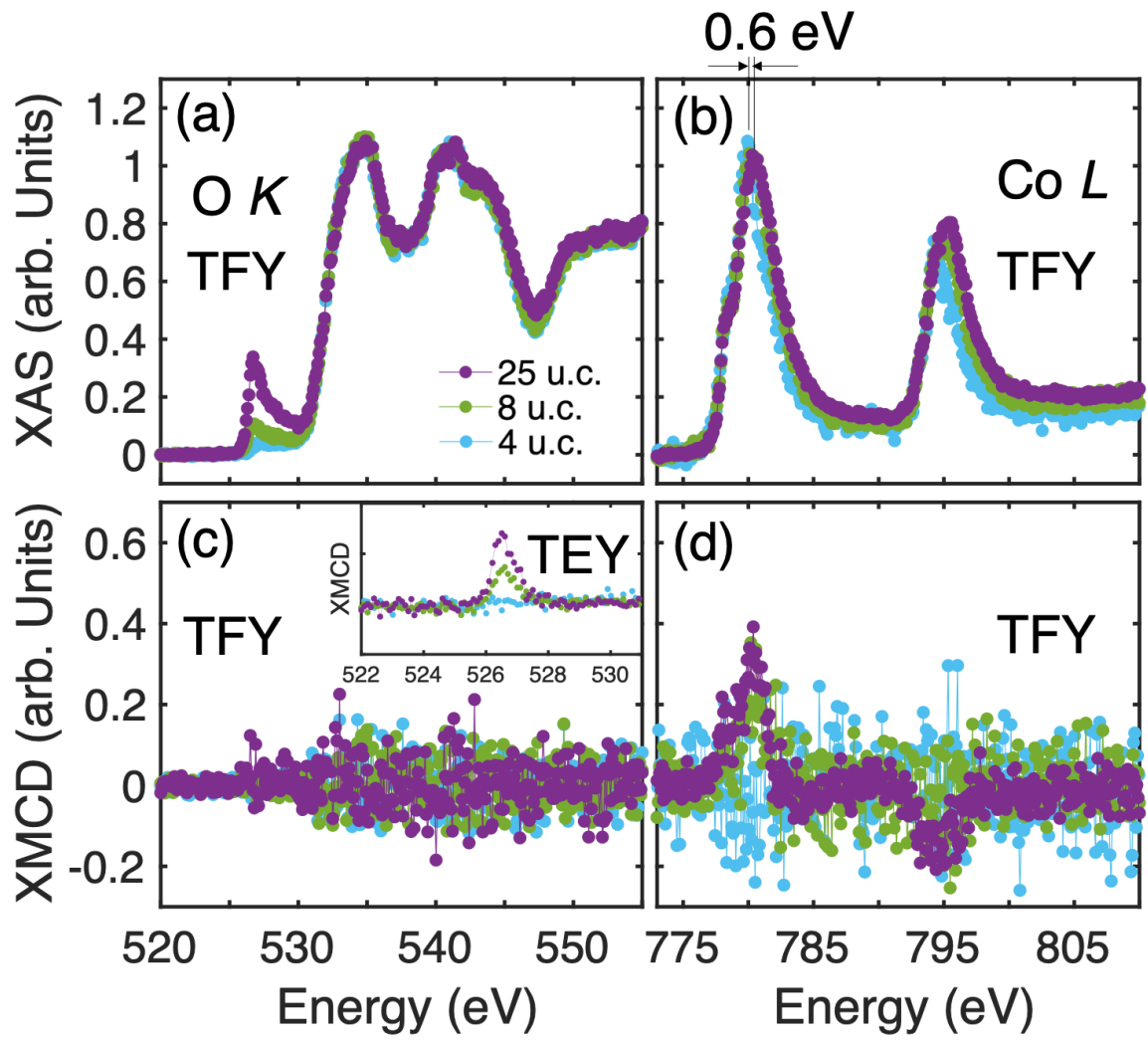

**Figure 1**

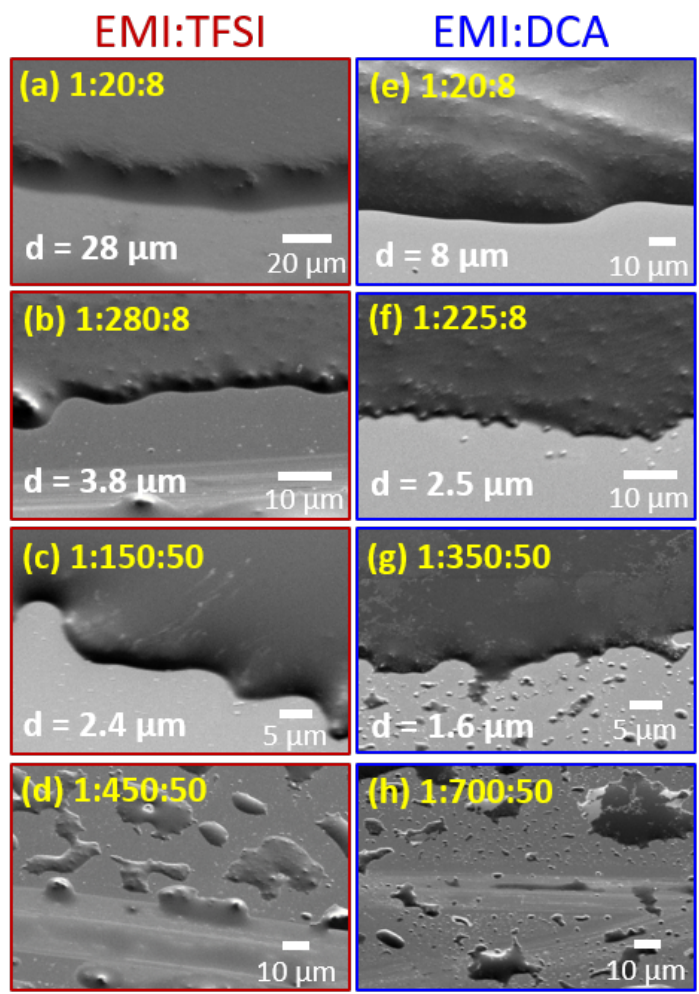

**Figure 2**

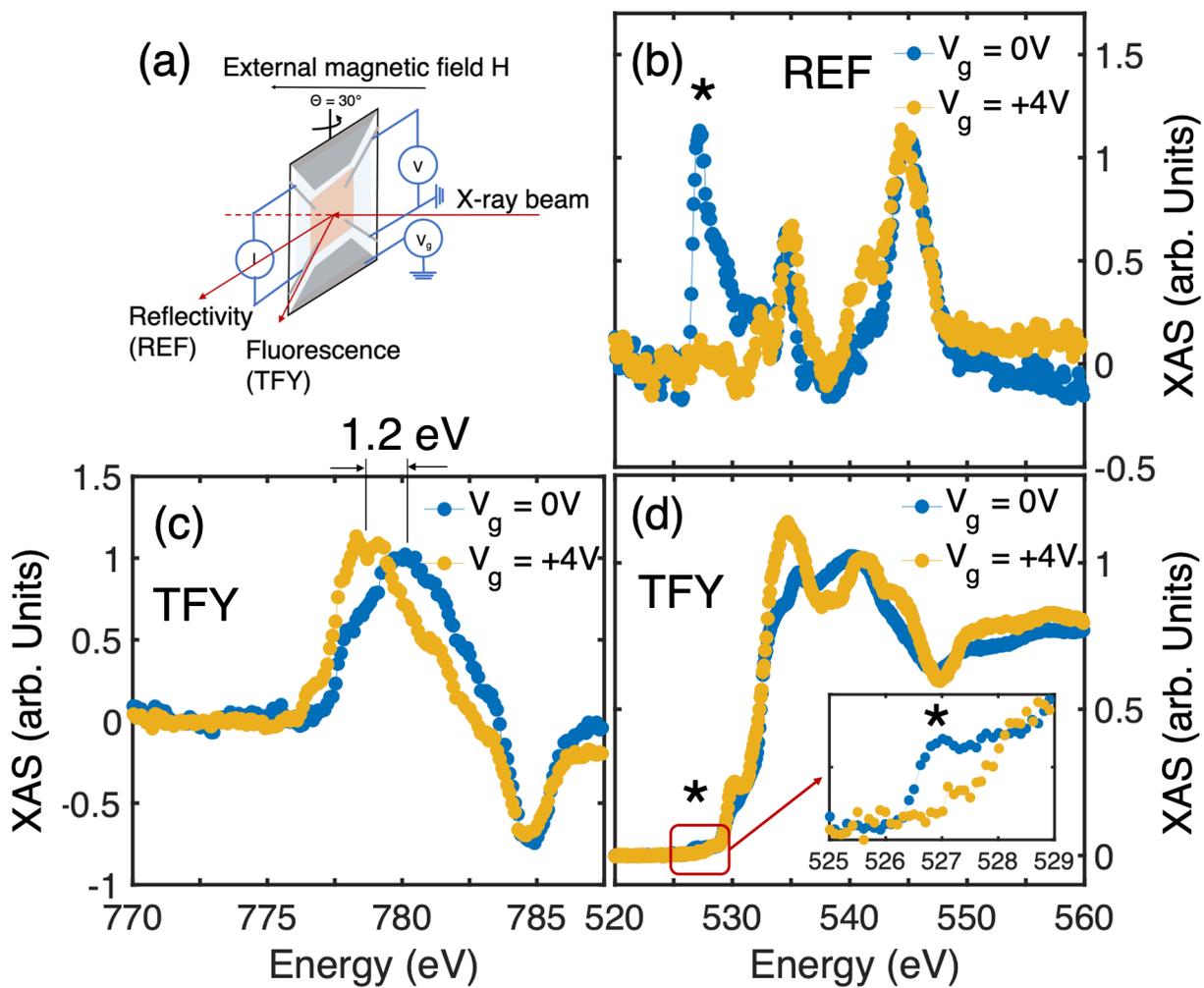

**Figure 3**

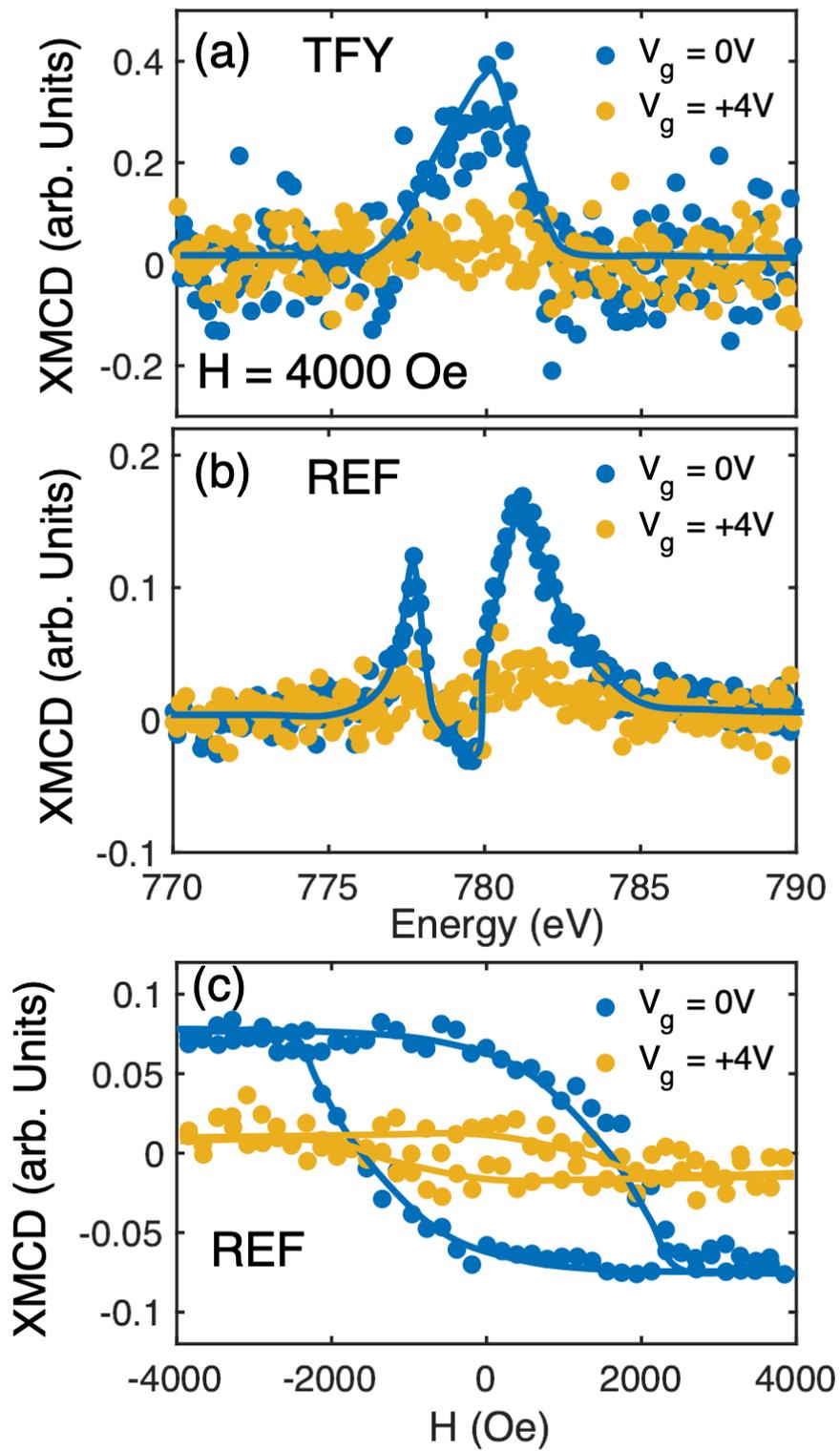

**Figure 4**